\documentclass[aps,prl,twocolumn]{revtex4}
\usepackage{graphicx}
\usepackage{amssymb}
\usepackage[dvips]{color}
\newcommand{\bfk}{{\bf k}}
\newcommand{\ek}{\epsilon_\bfk}
\newcommand{\cks}{c_{\bfk,\sigma}}
\newcommand{\cku}{c_{\bfk,\uparrow}}
\newcommand{\ckpd}{c_{\bfk',\downarrow}}
\newcommand{\cmkd}{c_{-\bfk,\downarrow}}
\newcommand{\bk}{b_\bfk}
\newcommand{\bz}{b_0}
\newcommand{\bfC}{{\bf C}}
\newcommand{\bfu}{{\bf u}}

\begin{document}
\title{Boson-like quantum dynamics of association in ultracold Fermi gases}
\author{I. Tikhonenkov and A. Vardi}
\affiliation{Department of Chemistry, Ben-Gurion University of the
Negev, P.O.B. 653, Beer-Sheva 84105, Israel} 

\begin{abstract}
We study the collective association dynamics of a
cold Fermi gas of $2N$ atoms in $M$ atomic modes into
a single molecular bosonic mode. The many-body fermionic 
problem for $2^M$ amplitudes is effectively reduced to a dynamical
system of $\min\{N,M\}+1$ amplitudes, making the solution no more complex 
than the solution of a two-mode Bose-Einstein condensate 
and allowing realistic calculations with up to $10^4$ particles. 
The many-body dynamics is shown to be formally 
similar to the dynamics of the bosonic system under the mapping
of boson particles to fermion holes, producing 
collective enhancement effects due to many-particle constructive
interference. Dissociation rates are shown to enhance as the 
number of particles whereas association rates are enhanced 
as the number of holes, leading to boson-like collective behavior.     

\end{abstract}


\maketitle
The role of Bose stimulation in the dynamics of coupled atomic
and molecular Bose-Einstein condensates (BECs) has been extensively
studied since the first theoretical proposals for the conversion of
atomic condensates into molecular BECs \cite{Timmermans99,vanAbeelen99,
Javanainen99,Heinzen00,Vardi01,Moore02}. It has been shown that the 
entire condensate will undergo large amplitude coherent oscillations
between atoms and molecules. The Bose-enhanced oscillation frequency 
is predicted in mean-field theory to scale as $\sqrt{N}g$ where 
$g$ is the 
single-particle atom-molecule coupling frequency and $N$ is the total 
number of condensate particles. Quantum-field effects slightly modify 
the effective frequency to $(\sqrt{N}/\ln N)g$ and introduce 
collapse and revival of the coherent oscillations due to interparticle 
entanglement \cite{Javanainen99,Vardi01}. Yet despite these modifications 
the collective behaviour remains significantly different than the
single pair dynamics, indicating the dramatic effect of quantum statistics
on the many-body dynamics.

Experimentally, the effort towards a molecular BEC and the realization
of the theoretical predictions, was made using the techniques of Raman 
photoassociation \cite{Wynar00} and 
Feshbach resonance magnetoassociation \cite{Donley01,Claussen01,Donley02}.
Unexpectedly, the first molecular condensates were produced using
the Feshbach method from nearly degenerate gases of {\it fermion} atoms 
\cite{Jochim03,Greiner03,Zwierlein03} which are more stable against
vibrational quenching than their bose-atom counterparts. Molecular BECs made of
boson constituent atoms came a close second \cite{Xu03,Mukaiyama04}. 

The possibility of coupling a nearly degenerate atomic Fermi gas with
a molecular BEC raises interesting questions about the nature of the
ensuing collective dynamics. At first site it appears that the dynamics
of associating fermions should be drastically different than the
association dynamics of bosons, since the former are subject to Pauli 
blocking as compared to the Bose stimulation affecting the time evolution
of bosons. However, it has been previously noted that collective 
effects in fermion systems can mimik bosonic stimulation in four-wave
mixing \cite{Moore01,Ketterle01}, where the boson-like behaviour is 
attained from various pathways adding up 
constructively \cite{Moore01}. Moreover, few-particle numerical results 
for the association of a Fermi-Bose mixture of atoms \cite{Dannenberg03}, 
indicate some similarity with the purely bosonic case, in that collective 
Rabi-like oscillations and Rapid adiabatic passage are observed in both cases.
Better understanding of this boson-like behaviour, allowing
for the simplification of calculations and illuminating its origin   
is highly desirable. 

In this work we study the association dynamics of an atomic fermion gas
into a degenerate molecular bose gas. We show that the many-fermion 
problem which is
apparently much more obstruse than the corresponding two-mode 
bosonic problem (or than the Fermi-Bose mixture case), can be 
effectively reduced to an $(N+1)\times(N+1)$ system of dynamical
equations, where $N$ is the total number of pairs (i.e. the sum 
of the molecule number and half the atom number). The
complexity of the system is thus identical to that of the dynamical  
equations obtained for two-mode bose association in its number-state
representation \cite{Javanainen99,Vardi01}, allowing for numerical 
calculations with up to $N=10^4$ particles with current computation power,
as opposed to characteristically $N=10$ particles in Ref. \cite{Dannenberg03}.

Furthermore, we show formally that in the limit of large $N$ and $M$ where the
number of fermion modes $M$ is equal to the number of particles $N$, the
structure of the fermionic system is a 'mirror image' of the
bosonic system, producing precisely the same dynamics for fermion dissociation
as was attained for boson association. Similarly the dynamics of
association into fermions reproduces the known results for  
boson dissociation, including its modulational instability \cite{Vardi01}. 
It is evident from our model that the enhancement effects come from adding up the 
various pathways connecting states with $N-n$ molecules and $n$ dissociated 
pairs with states containing one more (or one less) molecule and one less
(or one more) dissociated pair. Thus, the origin of the nonlinear collective
behaviour is found to be identical to the four-wave mixing case \cite{Moore01}.

\begin{figure}
\centering
\includegraphics[scale=0.47,angle=0]{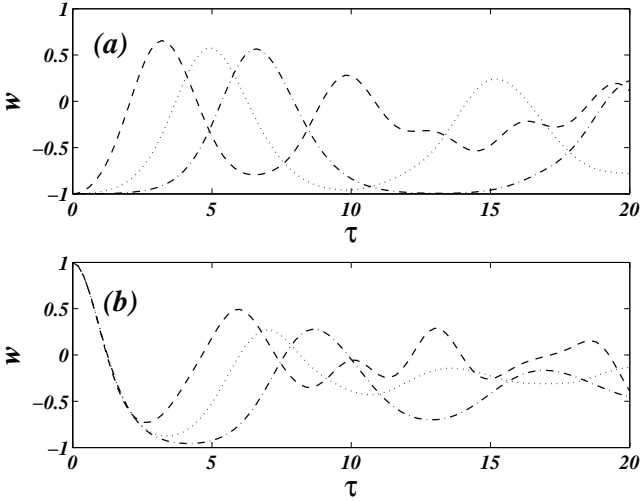}
\caption{Collective dynamics of association of an atomic Fermion gas into 
a molecular BEC (a) and dissociation of a molecular BEC into a 
degenerate Fermi gas of atoms (b). Particle pair numbers are
fixed to $N=5$ (dashed line), $N=50$ (dotted line), and $N=500$ (dash-dotted
line) pairs. The number of fermion modes is set equal to the number
of particles.} 
\end{figure}

We consider the association of a Fermi gas of atoms into Bose 
molecules. The interaction representation Hamiltonian reads
\begin{eqnarray}
H&=&\sum_{\bfk,\sigma} \ek\cks^\dag\cks+\sum_\bfk {\cal E}_\bfk \bk^\dag\bk\nonumber\\
~&~&+\left(g\sum_{\bf k} \cku\ckpd b_{\bfk+\bfk'}^\dag+H.c.\right)~,
\label{ham}
\end{eqnarray}
where $\ek=\hbar^2k^2/2m$ is the kinetic enery for an atom with mass $m$, 
${\cal E}_\bfk$ is the molecular energy containing both kinetic and binding 
conributions, $\cks$ are the annihilation operators for the atoms, obeying 
the usual Fermi anticommutation relation, $\bk$ are the molecular annihilation 
operators, commuting with $\cks$ and obeying bosonic commutation relations, 
and $g$ is the strength of the atom-molecule coupling. Since the pairing 
instability occurs predominantly for colliding atom pairs whose center of 
mass momentum is zero \cite{Mahan}, as confirmed by the formation of a 
molecular BEC in the experiments of Refs. \cite{Jochim03,Greiner03,Zwierlein03},
we use the single molecular mode approximation \cite{Javanainen04,Barankov04}, 
assuming that only singlet pairs 
of opposite momenta are associated, resulting in molecules in the lowest energy 
molecular mode. Moreover, since the Fermi energy is characteristically very
small compared with the atom-molecule coupling frequency, we can safely ignore 
the variation in $\ek$. This approximation is tantamount to assuming that
the motional timescale of the atoms is much longer than the timescale
of atom-molecule conversion \cite{Dannenberg03} and is verified by the 
nearly complete conversion of Fermi atoms to Bose molecules in Feshbach 
sweep experiments
\cite{Jochim03,Greiner03,Zwierlein03} (we note paranthetically that the variation
in $\ek$ can be easily incorporated into our model by proper renormalization
of the amplitudes). The resulting rotating wave Hamiltonian is
\begin{equation}
H=\Delta\left(\frac{1}{2}\sum_{\bfk,\sigma}\cks^\dag\cks
-\bz^\dag\bz\right) 
+\left[g\sum_{\bf k} \cku\cmkd\bz^\dag+H.c.\right]~,
\label{hamtwo} 
\end{equation}
where $\Delta=2\epsilon_{k_F}-{\cal E}_0$. In what follows we shall assume
that $\Delta=0$, i.e. we consider the case of resonant coupling between 
atoms and molecules, retaining only the interaction Hamiltonian. Since the 
Hamiltonian (\ref{hamtwo})
conserves the number of atoms, i.e. $2n+\sum_{\sigma,\bfk}n_{\sigma,\bfk}=2N$
where $n$ is the numeber of molecules, $n_{\bfk,\sigma}$ is the
number of atoms in state $\bfk,\sigma$ and $N$ is the (conserved) 
total number of pairs, we can fix $N$ and expand the quantum state 
of the system in pair number states:
\begin{equation}
|\psi\rangle=\sum_{n=0}^{\min\{N,M\}}\sum_{\{n_k\}} C_{N-n,n_1,\dots,n_M}(t)
|N-n,n_1,...,n_M\rangle
\label{psi}
\end{equation} 
where $k$ denotes some labeling of the $M$ available momentum states, 
$n_k=0,1$ are fermionic occupation numbers for the free
atom states (where $n_k=1$ denotes a singlet pair of particles occupying 
the momentum states $\bfk,-\bfk$) and the summation over $\{n_k\}$ for any given 
$n$ is over all nonordered combinations of $n$ atom pairs among
$M$ states, so that $\sum_{k=1}^M n_k=n$. 
The effective phase-space for the problem can thus be classified
into $\min\{N,M\}+1$ manifolds characterized by the number of dissociated
pairs $n=0,1,\dots,\min\{N,M\}$. For each $n$, the size of the corresponding
manifold is the number of orderings of $n$ pairs in $M$ states given 
by the binomial coefficient $\left(^M_n\right)$. It is immediately 
seen that the interaction Hamiltonian couples each state in the manifold 
with $n$ dissociated pairs to $M-n$ states having $n+1$ dissociated 
pairs and to $n$ states with $n-1$ free atom pairs. Substituting
the expansion (\ref{psi}) into the Schr\"odinger equation 
$i\hbar\partial_t|\psi\rangle=H|\psi\rangle$, with the Hamiltonian
(\ref{ham}), we obtain:
\begin{eqnarray}
i\dot{C}_{N-n,n_1,\dots,n_M}&=&g\left(\sqrt{N-(n-1)}\right.\\
~&~&\times\sum_{k=1}^M
\delta_{1,n_k}C_{N-(n-1),n_1,\dots,n_k-1,\dots,n_M}\nonumber\\
~&~&+\sqrt{N-n}\nonumber\\
~&~&\left.\times\sum_{k=1}^M\delta_{0,n_k}C_{N-(n+1),n_1,\dots,n_k+1,\dots,n_M}
\right)\nonumber
\end{eqnarray}
\label{eomo}
where $n=0,\dots,\min\{N,M\}$. 
These equations of motion can be put into the matrix form:
\begin{equation}
i\dot{\bfC}_n=g\left(\sqrt{N-n+1}D_n^{n-1}{\bfC}_{n-1}
+\sqrt{N-n}D_n^{n+1}{\bfC}_{n+1}\right)~,
\label{eomtwo}
\end{equation}
where $\bfC_n(t)$ is a column vector of all the $\left(^M_n\right)$ amplitudes 
corresponding to the possible arrangements of the $n$ atom pairs among the $M$ available 
fermion modes. The matrices 
$D_I^J$ are $\left(^M_I\right)\times\left(^M_J\right)$ dimensional and contain
only unit and zero elements. Their explicit form is determined by the 
ordering of elements in the vectors $\bfC_I$ and $\bfC_J$. 

The system 
(\ref{eomtwo}) contains equations for $2^M$ amplitudes, making effective 
dynamical calculations restricted to limited numbers of particle 
(characteristically no more than 20 particles with current computation
power). However, it is clear that the high symmetry of the system should
produce conservation laws that would allow its simplification. In fact, 
since all the states corresponding to various atom ordering within 
each $\left(^M_n\right)$ manifold of $N-n$
molecules are equivalent (their coupling strength to states with one more
or one less molecule are equal and each is coupled to the same number
of states above and below it), there are $\left(^M_n\right)-1$ constraints
in each such manifold, leaving effectively only $\min\{N,M\}+1$ free amplitudes.
Consequently, an enormous reduction in the dimensionality of the system
is attained by multiplication of Eq. (\ref{eomtwo}) by a 
row vector $\bfu_{n,M}$ of $\left(^M_n\right)$ unit elements (thus summing
over all amplitudes in the manifold of $n$ dissociated pairs)
and use of the identities:
\begin{equation}
\bfu_{n,M}D_n^{n+1}=(n+1)\bfu_{n+1,M}~,
\label{ido}
\end{equation}
\begin{equation}
\bfu_{n,M}D_n^{n-1}=(M-n+1)\bfu_{n-1,M}~.
\label{idt}
\end{equation}
The resulting equations of motion read
\begin{eqnarray}
\label{eomred}
i\dot\alpha_n&=&g\left(\sqrt{N-n+1}\sqrt{n(M-n+1)}\alpha_{n-1}\right.\nonumber\\
~&~&\left.+\sqrt{N-n}\sqrt{(M-n)(n+1)}\alpha_{n+1}\right)
\end{eqnarray}  
where
\begin{equation}
\alpha_n\equiv\frac{\bfu_{n,M}\bfC_n}{\sqrt{\left(^M_n\right)}}=
\frac{\sum_{\{n_k\}}C_{N-n,n_1\dots,n_M}}{\sqrt{\left(^M_n\right)}}
~,~\sum_{k=1}^M n_k=n,
\end{equation}
is the sum over all $\left(^M_n\right)$ amplitudes with $N-n$ molecules
and $n$ free atom pairs, corresponding to the various orderings of the
atoms, normalized by the number of permutations of $n$ atoms in $M$ 
states. The average number of molecules, provided that all amplitudes 
in the $n$-th manifold are equivalent (i.e. under the restriction 
that initial conditions do not break this equivalence), is calculated 
according to
\begin{eqnarray}
\langle\bz^\dag\bz\rangle&
=&\sum_{m=0}^{\min\{N,M\}} (N-n)\sum_{\{n_k\}}|C_{N-n,n_1\dots,n_M}|^2\nonumber\\
~&~&=\sum_{m=0}^{\min\{N,M\}} (N-n)|\alpha_n|^2
\end{eqnarray} 

\begin{figure}
\centering
\includegraphics[scale=0.47,angle=0]{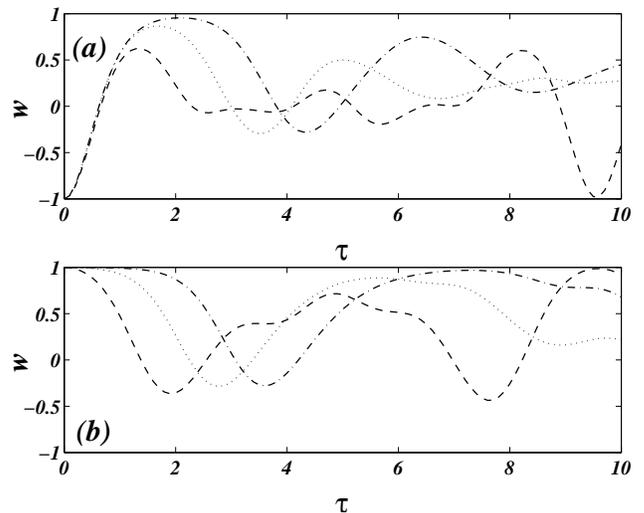}
\caption{Collective dynamics of association (a) and dissociation (b) 
in a two-mode atom-molecule BEC. Particle numbers are
fixed to $N=5$ (dashed line), $N=50$ (dotted line), and $N=500$ (dash-dotted
line) pairs. The number of fermion modes is set equal to the number
of particles.} 
\end{figure}

The size of the system (\ref{eomred}) is thus reduced
to $\min\{N,M\}+1$, exactly as would be obtained for the equivalent two mode bosonic
system, allowing numerical results with up to $10^4$ particles. The 
physical interpretation of Eqs. (\ref{eomred}) is quite
striking. It shows that in addition to the usual bosonic factors of 
$\sqrt{N-n+1}$ for association and $\sqrt{N-n}$ for dissociation, originating
from the molecular bosonic field operator $\bz$, the association
rate depends on the number of pairs in the initial state
(a trivial factor originating from having $n$ pairs which can 
associate) but is also enhanced by the number of {\it holes} in the target state.
Similarly, the rate of dissociation (i.e. the annihilation of a hole)
scales with the number of holes in the initial state but also with
the number of {\it particles} (or more precisely, particle-pairs)
in the target state.

The resulting dynamics should thus be surprisingly similar to the
dynamics of an atom-molecule condensate. It is illuminating to
compare Eq. (\ref{eomred}) to the $(N+1)\times(N+1)$ number-state 
representation of the Schr\"odinger equation for a two-mode
atom-molecule BEC \cite{Javanainen99,Vardi01}
\begin{eqnarray}
i\dot\beta_n&=&g\left(\sqrt{N-n+1}\sqrt{2n(2n-1)}\beta_{n-1}\right.\nonumber\\
~&~&\left.+\sqrt{N-n}\sqrt{(2n+2)(2n+1)}\beta_{n+1}\right),
\label{eombos}
\end{eqnarray}
where $n=0,\dots,N$ and the average molecule number is 
$\langle b^\dag b\rangle=\sum_{n=0}^N (N-n)|\beta_n|^2$.
Examination of Eq. (\ref{eombos}) shows that  both the 
association rate and the dissociation rate are enhanced 
as the number of atoms in the target state. Moreover, it
is evident that up to factors of two, the main difference
between the fermion equation (\ref{eomred}) and the 
boson equation (\ref{eombos}) is in the enhancement
factors where particle number terms $n$ in the boson 
case are replaced by hole number terms $M-n$ for fermions.
It is therefore clearly evident that when $N=M$ (i.e.
the number of pair states is equal to the number of pairs) 
the dynamics of fermion association (dissociation) should 
be qualitatively very similar to the dynamics of boson 
dissociation (association), up to a factor of two in
the pertinent rates. 

This observation is confirmed by numerical integration of 
Eq.(\ref{eomred}) and Eq.(\ref{eombos}) for
fermions and bosons respectively. In Fig. 1 and Fig. 2 
we plot the expectation value of the pair-number difference 
$w=2\langle \bz^\dag \bz\rangle/N-1$  as a function of
the rescaled time $\tau=\sqrt{2N}gt$. Initial conditions
had all free fermion atoms in (a) or all molecules in (b).  
The fermion results in Fig. 1 mirror the boson results 
of Fig. 2 (identical to Refs. \cite{Javanainen99,Vardi01}) 
in that the dynamics of fermion association 
closely resembles the dynamics of boson dissociation and 
vice versa. The similarities include the $\sqrt{N}/\ln N$
scaling of the atom-molecule oscillation frequency \cite{Vardi01} 
and the dynamical instability of the fermion atoms to molecule 
formation, noted previously in mean-field studies \cite{Javanainen04,Barankov04}. 
This instability turns 
out to be the mirror image of the familiar modulational instability of 
the molecular mode in the two-mode BEC dynamics \cite{Vardi01}. As expected,
a factor of two exists in the characteristic frequencies.

The mirror symmetry between fermion and boson association 
dynamics in the limit of large $N=M$ is in fact exact. To show
this we note that for large $N$ the boson equations of motion (\ref{eombos})
for the amplitudes $\beta_{N-m}$ may be written as 
\begin{eqnarray}
i\dot\beta_{N-n}&=&2g\left(\sqrt{n+1}(N-n)\beta_{N-n-1}\right.\nonumber\\
~&~&\left.+\sqrt{n}(N-n+1)\beta_{N-n+1}\right),
\end{eqnarray}
where we have neglected some terms of order $1\ll 2N$.
Mapping $\beta_{N-n}\rightarrow\beta_n$ we obtain that
\begin{eqnarray}
i\dot\beta_n&=&2g\left((N-n+1)\sqrt{n}\beta_{n-1}\right.\nonumber\\
~&~&\left.+(N-n)\sqrt{n+1}\beta_{n+1}\right),
\label{fbeq}
\end{eqnarray}
which is identical up to a factor of two, to the fermion
equation (\ref{eomred}) when $N=M$. This factor of two 
emanates from having a single atomic mode with $2n$ particles
in the bosonic problem, making it the analog of a degenerate 
optical parametric amplifier (OPA), as opposed to having $n$
particles each of spin $\uparrow$ and spin $\downarrow$ 
atoms in the fermion problem. The correspondence is thus even 
better with the nondegenerate OPA-like association of a
two species BEC 
\begin{equation}
H=g\left(a_1 a_2 b^\dag + b a_2^\dag a_1^\dag\right),
\label{ndb}
\end{equation} 
where $a_1,a_2$ and $b$ are all boson annihilation operators. 
The number-state representation of the Schr\"odinger equation
with the Hamiltonian (\ref{ndb}) maps {\it exactly}
to Eq. (\ref{fbeq}) under $\beta_{N-n}\rightarrow\beta_n$, without 
the multiplying factor of two and the restriction of large
$N$. The resulting dynamics of this boson problem at any 
$N$ is thus just the mirror image of Fig. 1.
 
The dynamics of the fermion amplitude corresponding to $n$ 
atom pairs (and $M-n$ hole pairs) thus maps {\it precisely} for $N=M$, 
to the  exact quantum dynamics of the two-species boson amplitude 
corresponding to $N-n$ atom pairs. At the limit of large $N$, 
the two-species boson dynamics reproduces the single atomic
mode boson dynamics with a factor of two in the characteristic
timescales. One can thus solve the fully quantum fermion problem
by simply solving the nondegenerate (two atomic species) two-mode 
BEC problem, and mapping boson amplitudes with $n$ particles to 
fermion amplitudes with $n$ holes, or at the limit of large $N$, solve the 
degenerate (single atomic specie) two-mode BEC problem, carry out the
same particle-hole mapping, and divide the timescale by two. 

To conclude, the apparently complex dynamics of an ultracold 
Fermi gas coupled to a molecular BEC can be greatly simplified
due to the inherent symmetry of the problem. We have obtained 
a system of dynamical equations for $N+1$ effective amplitudes
enabling an improvement of three orders of magnitude in
the total particle number of a realistic calculation. The
structure of this system provides great insight into the 
origin of boson-like collective behavior. It demonstrates
that dissociation is enhanced as the number of particles whereas 
association is enhanced as the number of holes, due to constructive
interference of various pathways between states with different
molecule number. Formal identity between fermion and boson dynamics
was shown to exist when $N=M$ and confirmed 
by numerical calculations. This equivalence can serve as
an important tool in further studies of nearly degenerate 
fermion-boson systems.      

\begin{acknowledgments}
We thank Ehoud Pazy for valuable discussions. This work
was supported in part by grants from the U.S.-Israel Binational Science
Foundation (grant No.~2002214), the Minerva Foundation
through a grant for a Minerva Junior Research Group, and the Israel
Science Foundation for a Center of Excellence (grant No.~8006/03).
\end{acknowledgments}

\end{document}